\documentclass[conference]{IEEEtran}
\IEEEoverridecommandlockouts
\usepackage{cite}
\usepackage{amsmath,amssymb,amsfonts}
\usepackage{algorithm}
\usepackage{graphicx}
\usepackage{textcomp}
\usepackage{xcolor}
\usepackage[utf8]{inputenc}
\usepackage{listings}
\usepackage{float}
\usepackage{subfigure}
\usepackage{url}
\usepackage{multirow}
\usepackage{mathptmx}
\usepackage{commath}
\usepackage{algpseudocode}
\usepackage{etoolbox}
\usepackage{varwidth}

\usepackage{tikz}
\usetikzlibrary{shapes.geometric, arrows}
\tikzstyle{startstop} = [rectangle, rounded corners, minimum width=2.5cm, minimum height=0.5cm,text centered, draw=black, fill=red!30]
\tikzstyle{io} = [trapezium, trapezium left angle=70, trapezium right angle=110, minimum width=2.3cm, minimum height=0.7cm, text centered, draw=black, fill=blue!30]
\tikzstyle{process} = [rectangle, minimum width=2.5cm, minimum height=0.8cm, text centered, text width=7cm, draw=black, fill=orange!30]
\tikzstyle{decision} = [diamond, minimum width=3cm, minimum height=1cm, text centered, draw=black, fill=green!30]
\tikzstyle{arrow} = [thick,->,>=stealth]

\def\BibTeX{{\rm B\kern-.05em{\sc i\kern-.025em b}\kern-.08em
    T\kern-.1667em\lower.7ex\hbox{E}\kern-.125emX}}
\begin{document}

%\title{Modeling the Performance of Multi-Threaded Applications using Reuse Distance Analysis}
%\title{Predicting Shared Cache Performance of OpenMP Applications on Multicores from the Memory Trace of a Sequential Run}
\title{Modeling Shared Cache Performance of OpenMP Programs using Reuse Distance}

\author
{
    \IEEEauthorblockN
    {
    %\vspace{-0.1in}
    Atanu Barai$^\star$, Gopinath Chennupati$^\dagger$, Nandakishore Santhi$^\dagger$, Abdel-Hameed A. Badawy$^{\star}$, Stephan Eidenbenz$^\dagger$\newline
    \vspace{5pt}
    }
    
    \IEEEauthorblockA
    {
    $^\star$ Klipsch School of Electrical and Computer Engineering, New Mexico State University, Las Cruces, NM, USA\\
    $^\dagger$ Information Sciences Group, Los Alamos National Laboratory, Los Alamos, NM, USA\\
    $^\star$\{atanu, badawy\}@nmsu.edu, $^\dagger$\{gchennupati, nsanthi, eidenben\}@lanl.gov\\
    }
%\vspace{-25pt}
}

\maketitle

\begin{abstract}
Performance modeling of parallel applications on multicore computers remains a challenge in computational co-design due to the complex design of multicore processors including private and shared memory hierarchies. We present a Scalable Analytical Shared Memory Model to predict the performance of parallel applications that runs on a multicore computer and shares the same level of cache in the hierarchy. This model uses a computationally efficient, probabilistic method to predict the reuse distance profiles, where reuse distance is a hardware architecture-independent measure of the patterns of virtual memory accesses. It relies on a stochastic, static basic block-level analysis of reuse profiles measured from the memory traces of applications ran sequentially on small instances rather than using a multi-threaded trace. The results indicate that the hit-rate predictions on the shared cache are accurate. %This analytical shared reuse profile for parallel programs can also be used to predict the effective latency and throughput of memory accesses, which in turn are used to predict the overall runtime of an application.
\end{abstract}


\section{Introduction}
With the emergence of Exascale computing, processors with more number of cores on a chip with complicated cache designs have become common. Such complicated designs come with a number of challenges~\cite{john:exascale}, such as efficient use of available compute power, modeling the performance of these complex caches, to name a few. Parallel applications that run on these multi-cores try to leverage this extensive computing power. One of the key factors that determine the performance of a parallel application on a multi-core processor is the availability of data to the cores. One way to measure the data availability of an application is through the cache utilization ability, which at the end impacts the performance.

Moreover, modern processors contain shared caches, which have a significant impact on the performance of an application in the form of data locality and inter-process communication. These factors are both complex to analyze and hardware dependent. Simulation, based on software/hardware co-design helps to better understand the behavior of applications and study the impact of the above factors on performance in a multi-core configuration. Co-design, in fact, helps to tune the performance of an application. Most of the efforts in co-design have focused on getting simulation data from cycle-accurate dynamic instrumentation tools~\cite{Moguls,Davis-max-cmp-thpt,Ekman-perf-pwr,Huh-des-sp}. However, these simulations require a large number of runs and require experimentation with a number of hardware configurations. Such configurations include variations in cache hierarchies, core counts and problem sizes, all of which contribute to increasing the complexity of design space exploration. Therefore, using cycle-accurate dynamic simulators to evaluate performance can be extremely challenging.

In analyzing the performance of a cache, \textit{Reuse Distance Analysis}~\cite{Mattson:RD:IBM} is one of the commonly used technique. Reuse distance is the number of unique memory references between two same consecutive accesses. For sequential programs, reuse analysis is architecture-independent, whereas for parallel programs that run on multi-core processors, reuse distance dependents on how the memory references of threads interact. Therefore, on multi-cores, \textit{Concurrent Reuse Distance} (CRD) profiles~\cite{Multicore:ding2009a} use a global stack to quantify reuse across thread-interleaved memory references, and thus accounts for data sharing and interaction between threads accessing shared caches. However, CRD profiles are unscalable as the core count increases, the thread interactions increase, thereby the memory traces get large, which significantly changes the CRD profiles. 

In this paper, we introduce a scalable reuse distance-based shared memory model in order to estimate the shared cache hit rates for different applications. We use a translator based on the {\em Rose} compiler~\cite{Rose_Compiler:Liao} to get the threaded version of a parallel code written in OpenMP. We develop a compiler-driven technique to identify the basic blocks of the threaded programs in measuring the exact probabilities of executing a given basic block of a program. We explore through different interleaving strategies of execution in order to mimic the behavior of multi-threaded programs on shared memory processors. In fact, these strategies are carried-out at the LLVM basic block~\cite{Lattner:LLVM}. The memory references of the labeled trace generated from the sequential run of the program and apply an analytical probabilistic method to measure reuse profiles of applications. Using these profiles, we measure cache hit-rates of the applications.
We evaluate our approach on two benchmarks Breadth-First Search (BFS) and Matrix Multiplication (MatMul) on two processors an Intel Core I7 and an Intel Xeon. We compare our results with the actual hit-rates calculated from the memory trace generated using Valgrind~\cite{valgrind} Lackey tool. The results show that the model predicted cache hit-rates are similar to that of the actual hit-rates.

\section{Background}
%\subsection{Performance Modeling}
\subsection{Execution of Parallel Application: Fork Join Model}
OpenMP uses fork-join model for parallel execution of a program. The program begins as a sequential application with master thread. When the first parallel region construct is encountered, the master thread splits itself (forks) into a team of identical parallel threads. The forked traces have access to all the variables from the master thread so those are shared variables. They also have private variables of their own and can identify themselves with unique thread number. When the team threads complete executing all the statements in the parallel region, they synchronize and terminate (join), leaving only the master thread. It's also possible to have nested parallelism where one of the team threads can split itself.

\subsection{Reuse Distance Analysis}
\begin{figure}[htbp]
    \centering
    \begin{tabular}{cc}
    \multicolumn{2}{c}
    {
        \centering
        \begin{tabular}{|c|cccccccc|}
           % Time & 1 & 2 & 3 & 4 & 5 & 6 & 7 & 8 \\
            \hline
            Address & a & b & a & c & b & d & d & a\\
            \hline
            {\text { RD }} & {$\infty$} & {$\infty$} & {1} & {$\infty$} & {2} & {$\infty$} & {0} & {3}\\
            \hline
        \end{tabular}
    } \\
   \if 0
    \\
    
        {
            \setlength{\tabcolsep}{2pt}
            \begin{tabular}{c|cccc|c}
                Distance &  0 & 1 & 2 & 3 & -1\\
                \hline
                Frequency & 1 & 1 & 1 & 1 & 4 \\
                \hline
                P(RD) & {1\(/ 8\)} & {1\(/ 8\)} & {1\(/ 8\)} & {1\(/ 8\)} & {4\(/ 8\)}
            \end{tabular}
        } 
        & 
        \includegraphics[width=30mm,scale=0.15]{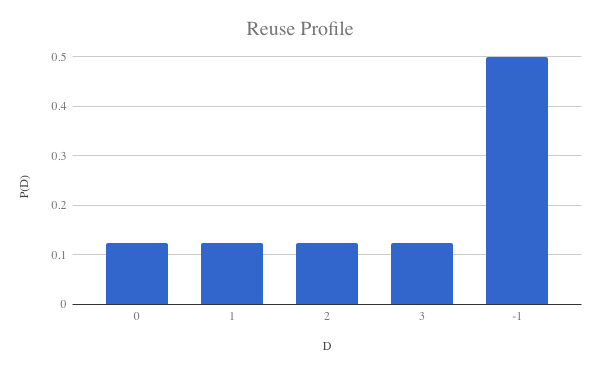}
    \fi
    \end{tabular}{}

    % \qquad
    
    % \begin{tabular}{c|cccc|c}
    %     Distance &  0 & 1 & 2 & 3 & $\infty$\\
    %     \hline
    %     Frequency & 1 & 1 & 1 & 1 & 4 \\
    %     \hline
    %     P(RD) & {1\(/ 8\)} & {1\(/ 8\)} & {1\(/ 8\)} & {1\(/ 8\)} & {4\(/ 8\)}
    % \end{tabular}
    % \newline
    % \textcolor{red}{compulsory miss = 50\%}
    
    \caption{Reuse Distance Example}
    \label{fig:RD_theory}
\end{figure}

Reuse distance (D) of a memory address which is also known as LRU stack distance is the number of unique memory references between two consecutive reference to the same address. Note that, when a memory address is referenced for the first time, the reuse distance, D is $\infty$. Reuse profile is the histogram of reuse distances for all memory references of a program. Reuse distance analysis measures the locality~\cite{locality:Ding:2003:PWL,locality:Zhong:2009:PLA} of an application, which can be used to predict cache performance of that application~\cite{performance:Beyls:RD:2001,performance:Sen:2013:ROM,performance:CaBetacaval:2003:ECM} and make cache management policy decisions~\cite{C.Management:Duong:2012:ICM}. For a fully associative cache with capacity C, the reuse distance of a memory reference will always trigger a cache miss, if D $\geq$ C. Fig.~\ref{fig:RD_theory} shows the reuse distance calculation for a sample trace. In the example, $50\%$ of memory references will cause a compulsory cache miss. If we consider that cache size is 3 then 13\% of all memory references will cause a capacity cache miss.

Reuse distance analysis is powerful in the sense that it is architecture-independent for sequential applications. The same reuse profile can be used to measure the performance of different cache sizes. This saves a significant amount of time in cache hit-rate analysis as we don't have to collect memory traces for different cache configurations. A number of attempts~\cite{Ding:2001:RDA,Berg:SS,Steen:UGhent} demonstrated the use of memory traces for reuse profile calculations. These approaches use binary instrumentation tools to collect memory traces. The memory traces used in most of these attempts are large and time-consuming to process, thereby unscalable. However, recent attempts in~\cite{ppt-amm,chennupati:pmbs,chennupati:pads,Shen:2007:LAU} demonstrated analytical models that scale with a small input run of a program.

\subsection{Multicore Reuse Distances}
%Previous multicore RD research has revolved around developing techniques for acquiring profiles dynamically and verifying accuracy which is not scalable. 
Most of the multicore processors contain both shared and private caches. Although the locality of references of a parallel program in a multicore processor is somewhat architecture-specific, it largely depends on the characteristics of the application itself. The corresponding thread of a core accesses the private cache while the shared cache is accessed through all the cores. Two separate reuse profiles, \textit{Concurrent} and \textit{Private-stack} reuse profiles (CRD and PRD) are used to model shared and private caches~\cite{Jiang:RD-Applicable-on-chip}. To measure concurrent reuse profile, we can interleave memory references from all cores on a single LRU stack. This interleaving causes different types of interaction: \textit{dilation, overlap, and interception}~\cite{Wu-multicore-journal}. Figure~\ref{fig:CRD_theory} shows the memory references from two cores. For access of \textbf{a} at time 4 CRD is 2 where it's D is 1. Here CRD is larger than D which shows \textit{dilation}. On the other hand data sharing reduces dilation. For the memory reference of \textbf{a} at time 8 CRD is 2 although there are 3 memory references between two consecutive memory references at time 4 and 8. This shows \textit{overlapping} as \textbf{d} is accessed by both cores inside reuse interval of \textbf{a}. Again for the reference \textbf{b} at time 9, the reused data itself is shared. So its CRD is 2 which is less than its D.

\begin{figure}[htbp]
    \centering
    \begin{tabular}{cc}
    \multicolumn{2}{c}
    {
        \centering
        \begin{tabular}{c|cccccccccc}
            % \hline
            Time       & 1 & 2 & 3 & 4 &  5 &  6 & 7 & 8 & 9 & 10\\
            \hline
            Core $C_1$ & a &   & b & a &  e &  &   & d & a & b\\
            \hline
            Core $C_2$ &   & c &   &   &   & d & b &   &   &  \\
            % \hline
        \end{tabular}
    } \\
    \end{tabular}
    \caption{Concurrent Reuse Distance Example}
    \label{fig:CRD_theory}
\end{figure}

Several recent works have focused on CRD profile and performance prediction of the shared cache~\cite{Multicore:Performance_metrics:Ding2014,Multicore:Modeling_Superscalar_Memory-Level_Parallelism,Multicore:Modeling_CMP_Cache_Capacity:Shi,Multicore:Miss_Rate_Prediction:Zhong,Multicore:Formalizing_Data_Locality:Ceballos}. Recently researchers attempted to use analytical model and sampling to speed up the performance prediction~\cite{Jiang:RD-Applicable-on-chip,Multicore_Reuse_Analytical:Jasmine,Schuff:2010:AMR:1854273.1854286,multicore:Fast_and_Accurate_Exploration:Maeda,multicore:stat_multiprocessor_cache:Berg,Multicore-Aware-Derek}. All these models require trace collection from parallel execution of an application for different number of threads. On the other hand, our model collects trace once from the sequential run of the application. From that trace, we predict shared cache performance for a different number of threads. This makes our model highly scalable with core counts.

\section{Scalable Analytical Shared Memory Model}
The scalable analytical shared memory model is a parameterized model for performance prediction of parallel codes. We leverage reuse distance analysis to determine multicore reuse profile of a parallel program that runs on multiple cores. The reuse profiles are later used to determine the hit-rates at different cache hierarchies. %Using these reuse profiles, we estimate the availability of program data from the main memory to the processor via different cache levels.
\if 0
\begin{figure}[htp]\footnotesize
	\centering
\begin{tikzpicture}[node distance=1.1cm]

% \node (start) [startstop] {Start};
\   (in1) [io] {Input Program};
\node (preproc1) [process, below of=in1] {Convert the OpenMP code to threaded code}
\node (preproc2) [process, below of=preproc1] {Add labels to the threaded code to identify references of shared variables in the memory trace}
\node (pro1) [process, below of=preproc2] {Generate a basic block labeled memory trace with a smaller input size of the program ran sequentially};
\node (pro2) [process, below of=pro1] {Estimate the analytical reuse profile for shared memory by mimicking for the multicore trace};
% \node (pro3) [process, below of=pro2] {Measure the effective latency, bandwidth};
\node (pro3) [process, below of=pro2] {Measure the cache hit rates};
\node (out1) [io, below of=pro3] {Runtime of a program};
% \node (stop) [startstop, below of=out1] {Stop};

% \draw [arrow] (start) -- (in1);

\draw [arrow] (in1) -- (preproc1);
\draw [arrow] (preproc1) -- (preproc2);
\draw [arrow] (preproc2) -- (pro1);
\draw [arrow] (pro1) -- (pro2);
\draw [arrow] (pro2) -- (pro3);
\draw [arrow] (pro3) -- (out1);
% \draw [arrow] (out1) -- (stop);
\end{tikzpicture}
	\caption{Different steps in analytical shared memory model}\label{fig:flowchart}
\end{figure}
\fi
\definecolor{codegreen}{rgb}{0,0.6,0}
\definecolor{codegray}{rgb}{0.5,0.5,0.5}
\definecolor{codepurple}{rgb}{0.58,0,0.82}
\definecolor{backcolour}{rgb}{0.95,0.95,0.92}
 
\lstdefinestyle{mystyle}{
    backgroundcolor=\color{backcolour},   
    commentstyle=\color{codegreen},
    keywordstyle=\color{magenta},
    numberstyle=\tiny\color{codegray},
    stringstyle=\color{codepurple},
    basicstyle=\ttfamily\footnotesize,
    breakatwhitespace=false,         
    breaklines=true,                 
    captionpos=b,                    
    keepspaces=true,                 
    numbers=left,                    
    numbersep=4pt,                  
    showspaces=false,                
    showstringspaces=false,
    showtabs=false, frame=single,                  
    tabsize=2
}

%\subsection{Preprocess Source Code}
\subsection{Source Code Translation}
In the first step, we convert the OpenMP application to an intermediate threaded code using OpenMP translator in Rose~\cite{Rose_Compiler:Liao} compiler. In the translation process, the parallel sections of the original code are transformed into intermediate threaded code. The threaded version of the code contains XOMP wrapper functions (generated from the Rose compiler), that call GNU OpenMP (GOMP) (when compiled with GCC) library functions. The private variables of the parallel sections are translated as local variables in the corresponding threaded version of the code. Each thread under execution runs the XOMP wrapper functions, where each thread allocates memory for the local variables. For the shared variables, the functions in the threaded version of the code receive a structure of pointers as a parameter. At the beginning of these functions, all the members of those structures are assigned to locally declared pointers. We create separate \emph{labels} for these shared parts of the code, which is where the assignments happen so that the memory trace of the shared variables of the code are grouped in the corresponding basic block labels (described in the next section). %Fig.~\ref{fig:simple_rose_code} shows the intermediate code for the simple OpenMP code shown in Fig.~\ref{fig:simple_openmp}.
\subsection{Shared Memory Trace Generation}
In the second step, we generate LLVM basic block labeled memory trace of the translated threaded program. The LLVM IR of the source code consists of basic blocks, each of which contains a single entry and exit points. In producing the trace, we execute the translated code sequentially with smaller input size of a program. We use LLVM based instrumentation facilities to generate the basic block labeled memory trace of the translated program by sequential execution. In this memory trace the \textit{i\textsuperscript{th}} basic block\textit{(BB\textsubscript{i})} of the labeled trace contains all the memory addresses that are accessed as a result of executing the corresponding straight-line code of \textit{(BB\textsubscript{i})}. For each shared variable, marked with a {\em label}, we gather the corresponding memory references of those shared sections. Using this memory trace resulted from a sequential execution, we mimic the behavior of the parallel program and generate the private memory trace on each thread under execution.

As OpenMP works in fork-join model, the parallel section of the OpenMP code is executed at the same time on different cores. Each core has its own copy of the parallel section of the code. Note that the master thread execute the sequential part of the code along with the corresponding parallel section of the code. We mimic this behavior by making copies of the memory references of each basic block of the parallel sections. Our mimicking strategy tries to replicate the memory trace of an OpenMP program on multiple cores. For example, if the parallel program is using 4 cores, then we take four copies of a basic block, we then add an {\em offset} to the memory addresses for each of the cores under execution. The basic blocks that we select obviously belong to the parallel region of the code. The offset is carried out on all the memory references of all the basic blocks of a parallel region except for the memory references of the shared variables. This mimicking strategy helps to show that the memory references belong to different cores.

We choose the offset in a way such that the mimicked memory references do not match with the original memory references that are produced in the sequential execution. The original OpenMP execution contains different scheduling strategies to execute the parallel sections of the code. Recording memory traces for such scheduling strategies are cumbersome and time and memory inefficient. Therefore, our model in this paper tries to generate a trace that looks similar to the OpenMP scheduled traces. Here, we use the above recorded sequential trace with offsets in order to mimic the interleaving of threads. Our interleaving strategy distributes the corresponding memory threads equally among multiple threads under execution, which is similar to following static scheduling of OpenMP. We can also distribute the iterations to the cores according to an adaptive chunk sizes. We can explore through various interleaving and scheduling strategies, which is beyond the scope of this paper and we reserve it for future work. In this way, we find the private memory trace for each core under execution.

For shared memory, we take the {\em labeled} shared memory references from the sequential trace. These memory references can be found in the private trace from above, whereas we are not adding any offset to these variables. However, these references have the same memory address across multiple cores in the mimicked trace. These memory references are interleaved in the same way as the private variables. Similar traces can be generated with binary instrumentation tools such as Valgrind~\cite{valgrind} and Pin~\cite{pin-tool}, however, we use LLVM based tool to leverage the advantage of basic blocks of a program. Valgrind Lackey tool runs the multi-threaded program sequentially per thread, where the interleaving of the threads is left to the operating system, therefore the resultant memory trace happens to be a multi-threaded trace. On the other hand, with Pin, one has to produce a sequential trace and propose interleaving strategies. Nonetheless, we can not derive a basic block labeled trace from Pin as opposed to LLVM instrumentation. Once we have the memory trace that mimics the multi-core execution, we estimate the reuse distances for each reference in the trace.

\begin{figure*}[htp]
    \centering
    \begin{tabular}{p{0.50\textwidth}p{0.50\textwidth}}
    \begin{minipage}{0.47\textwidth}
      \centering
      \includegraphics[width=0.98\linewidth]{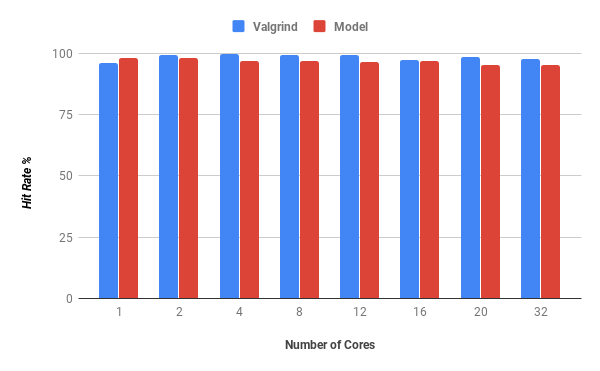} \vspace{-1em}
      \caption{Compare hit-rates between Valgrind and our model for BFS application on different number of cores at 6MB cache}\label{fig:6MB_BFS}
    \end{minipage}&
    \begin{minipage}{.47\textwidth}
      \centering
      \includegraphics[width=.98\linewidth]{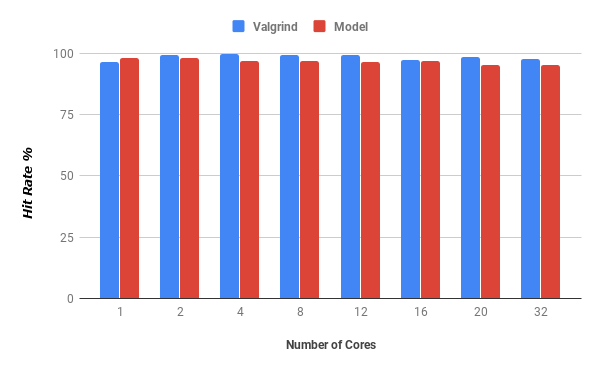} \vspace{-1em}
      \caption{Compare hit-rates between Valgrind and our model for BFS application on different number of cores at 25MB cache}\label{fig:25MB_bfs}
    \end{minipage}
    \end{tabular}
\end{figure*}

\begin{figure*}[htp]
    \centering
    \begin{tabular}{p{0.50\textwidth}p{0.50\textwidth}}
    \begin{minipage}{0.47\textwidth}
      \centering
      \includegraphics[width=0.98\linewidth]{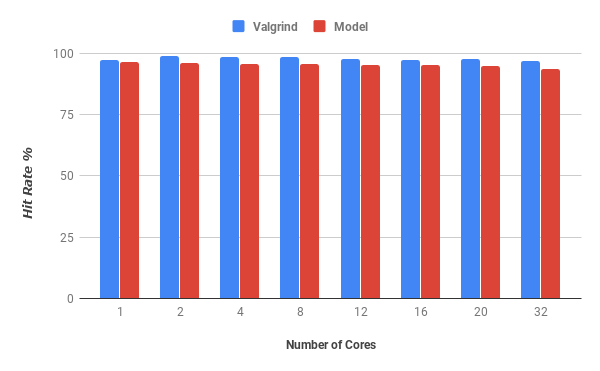} \vspace{-1em}
      \caption{Compare hit-rates between Valgrind and our model for MatMul application on different number of cores at 6MB cache}\label{fig:6MB_mm}
    \end{minipage}&
    \begin{minipage}{.47\textwidth}
      \centering
      \includegraphics[width=.98\linewidth]{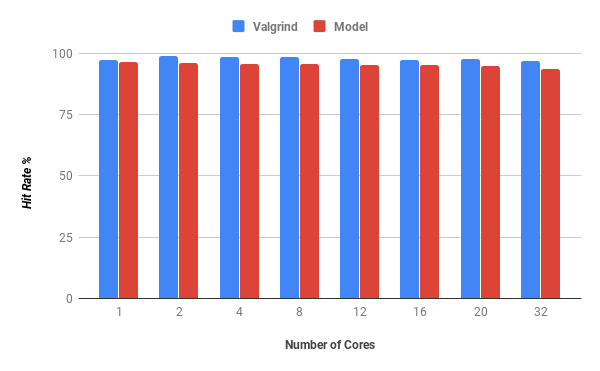} \vspace{-1em}
      \caption{Compare hit-rates between Valgrind and our model for MatMul application on different number of cores at 25MB cache}\label{fig:25MB_mm}
    \end{minipage}
    \end{tabular}
\end{figure*}

\subsection{Probabilistic Reuse Profile Estimation}
In our next step, we analytically estimate the probabilistic concurrent reuse profile of the program \textit{(Pr(D))} from our mimicked share memory trace. The conventional methods of measuring the reuse profile are expensive due to large memory traces. We use a technique described in~\cite{ppt-amm,chennupati:pmbs} which produces reuse distances at smaller input sizes of a program and from those reuse distances, we estimate reuse profiles at larger input sets. We estimate the reuse profile of a program using Eq.~\ref{eq:reuse_profile_program}.

\begin{equation}\label{eq:reuse_profile_program}
\operatorname{Pr}(D)=\sum_{i=0}^{n(B B)} P\left(B B_{i}\right) \times P\left(D | B B_{i}\right)
\end{equation}
\noindent where \textit{n(BB)} is the number of basic blocks, \textit{$P(BB_i)$} is the apriori probability of executing a basic block and \textit{$P(D|BB_i)$} is the conditional probability of executing a basic block.

\subsection{Hit Rate Estimation}
With the probabilistic reuse profiles (Pr(D)), we measure the shared cache hit-rates using an analytical memory model, a stack distance based cache model (SDCM)~\cite{brehob:analytical}. Eq.~\ref{eq:phd} shows how to measure the hit-rate at a given reuse distance ($P(h\mid D)$).
\begin{equation}\label{eq:phd}
P(h\mid D) =  \sum_{a=0}^{A-1}\binom{D}{a}\biggl(\dfrac{A}{B}\biggr)^a\biggl(\dfrac{B-A}{B}\biggr)^{(D-a)}
\end{equation}
\mbox{where} {\em D} is the reuse distance, {\em A} is the associativity of the cache and {\em B} is cache size in terms of number of blocks (which is cache size over cache line size).

\section{Experimental Results}
We evaluate the proposed model on two different CPU architectures with two benchmarks. The two architectures are Intel Core-I7 and Intel Xeon processors while the benchmarks are, breadth first search (BFS) and matrix multiplication (MatMul). The shared cache ($L_3$) sizes of both the architectures are $6 MB$ and $25 MB$ respectively. In order to validate, we compare the model predicted shared cache hit-rates with that of the hit-rates from Valgrind. Note that, we use Valgrind Lackey tool to get the memory trace of the benchmarks, compute reuse profiles, from which we measure the hit rates. Since the Lackey tool instruments the binary of the application at runtime, we generate the memory trace by defining the number of threads using $OMP\_NUM\_THREADS$ environment variable. We run the experiments for varying number of core counts, 1, 2, 4, 8, 12, 16, 20, 32. In each of these experiments we run Valgrind to record the traces and the reuse profiles, while our model runs the programs on one core and mimics the execution for other core counts.  %We use two cache configurations (6MB and 25MB cache size, both with associativity 20 and line size 64) and two applications (MatMul and BFS) to evaluate our results. 

Fig.~\ref{fig:6MB_BFS} and Fig.~\ref{fig:25MB_bfs} show the comparison hit-rates between our model and Valgrind Lackey tool for two different $L_3$ caches on BFS application, at different number of processors for an input size of 100 nodes. Fig.~\ref{fig:6MB_mm} and Fig.~\ref{fig:25MB_mm} show a similar comparison for MatMul with matrix sizes of 62x15 and 15x7. On an Intel Core-I7, the average hit-rates of BFS are 98.49\% for Valgrind and 96.77\% for our model, while for MatMul 97.83\% and 95.31\%, respectively. Similarly, on an Intel Xeon, the hit-rates are 98.51\% and 96.77\% for BFS and 97.90\% and 95.31\% for MatMul, respectively. On both the benchmarks, for both the cache configurations, the results show that our model predicts the shared memory hit-rates accurately. %The results show that our model can produce memory trace and thus predict cache hit-rates almost accurately like the Valgrind lackey tool. The result shows that our model can predict the shared cache hit rate with 97.61\% accuracy.

%Table~\ref{tab:runtime_comparison} shows the comparison of time required to collect memory trace using Valgrind on an Intel Xeon E5 processor for different threads and our model for matrix multiplication with matrix sizes 62x15 and 15x7.

\if 0
\begin{table}[htp]
    \centering
    \caption{Comparison of run-times to generate memory trace}
    \begin{tabular}{c|cccccccc}
        No. of Cores & 1 & 2 & 4 & 8 & 12 & 16 & 20 & 32 \\
        \hline
        Valgrind (Seconds) & 13 & 13 & 10 & 10 & 15 & 14 & 20 & 32\\
         \hline
        Model (Seconds)& 3 & - & - & - & - & - & - & - \\
         \hline
    \end{tabular}
    \label{tab:runtime_comparison}
\end{table}
\fi
%\section*{Acknowledgment}

\section{Conclusion}
Reuse distance analysis has been a valuable tool for application locality prediction, cache modeling, and performance prediction. This paper extends reuse distance analysis to the parallel application domain by accounting for inter-thread interactions for shared caches in a static way. It statically predicts the memory trace of a parallel application on shared cache from the memory trace of sequential execution of the code. This makes the method very scalar with core counts and cache sizes. The results show that our model is very accurate for the shared cache hit-rate prediction. Furthermore, the model takes the cache configuration parameters as input which makes it suitable for design space exploration and cache sensitivity analysis. 

In the future, we will extend the model to predict the hit-rates on private caches such as $L_1$ and/or $L_2$. Furthermore, We will explore various scheduling strategies of OpenMP with different interleaving strategies.
%PPT~\cite{ppt}

%\newpage
%\newpage
%\large{Other Results}
\if 0
\begin{figure}
    \centering
    \includegraphics[width=\linewidth]{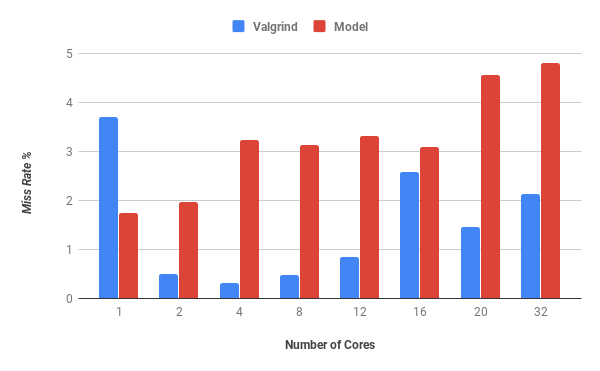}
    \caption{Comparison of miss rates on 6MB cache between Valgrind generated trace and trace from our model for BFS application on different number if cores}
    \label{fig:Miss_6MB_BFS}
\end{figure}
\begin{figure}
    \includegraphics[width=\linewidth]{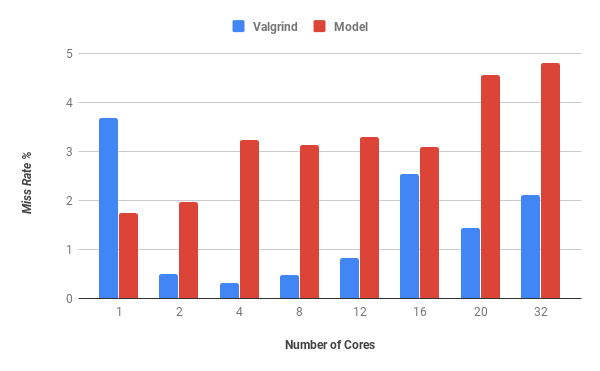}
    \caption{Comparison of miss rates on 25MB cache between Valgrind generated trace and trace from our model for BFS application on different number if cores}
    \label{fig:Miss_25MB_bfs}
\end{figure}
\begin{figure}
    \includegraphics[width=\linewidth]{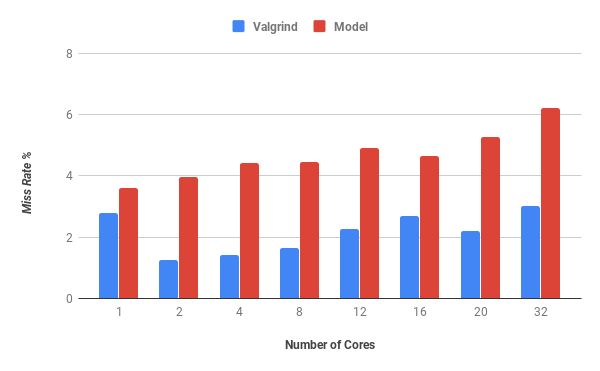}
    \caption{Comparison of miss rates on 25 MB cache between Valgrind generated trace and trace from our model for Matrix Multiplication application on different number if cores}
    \label{fig:Miss_6MB_mm}
\end{figure}
\begin{figure}
    \includegraphics[width=\linewidth]{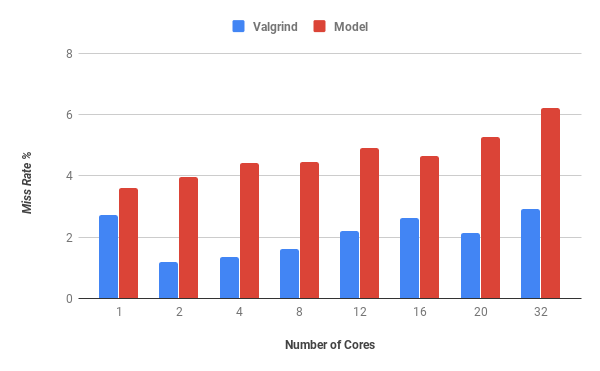}
    \caption{Comparison of miss rates on 25 MB cache between Valgrind generated trace and trace from our model for Matrix Multiplication application on different number if cores}
    \label{fig:Miss_25MB_mm}
\end{figure}
\fi

%\section*{Acknowledgement}

\bibliographystyle{IEEEtran}
\bibliography{references}

\end{document}